\documentclass[runningheads]{svmult}
\usepackage{makeidx}   % allows index generation
\usepackage{graphicx}  % standard LaTeX graphics tool
\usepackage{subeqnar}  % subnumbers individual equations
\usepackage{multicol}  % used for the two-column index
\usepackage{cropmark} % cropmarks for pages without
\usepackage{physprbb}  % centered layout of diverse elements, etc.
\begin{document}
\title*{The Amount of Dark Matter in Spiral Galaxies}
\toctitle{The Amount of Dark Matter in Spiral Galaxies}
% allows explicit linebreak for the table of content
%
%
\titlerunning{The Amount of Dark Matter in Spiral Galaxies}
% allows abbreviation of title, if the full title is too long
% to fit in the running head
%
\author{Burkhard Fuchs}
%
%\authorrunning{Ivar Ekeland et al.}
% if there are more than two authors,
% please abbreviate author list for running head
%
%
\institute{Astronomisches Rechen--Institut, M\"onchhofstr.~12-14,
69120 Heidelberg, Germany}

\maketitle              % typesets the title of the contribution

\begin{abstract}
The `maximum' disk hypothesis of galactic disks imbedded in dark matter halos 
is examined. First, decompositions of the rotation curves of NGC\,2613,
3198, 6503, and 7184 are analyzed. For these galaxies the radial velocity
dispersions of the stars have been measured. If the
parameters of the decompositions are chosen according to the `maximum' disk
hypothesis, the Toomre $Q$ stability parameter is sytematically less than one,
which is a strong argument against the `maximum' disk hypothesis. Next, density
wave theory arguments are used to describe the morphology of the spiral arms of
NGC\,3223, 157, and 7083. It is shown that the `maximum' disk hypothesis
is not consistent with the observed morphologies of the galaxies.
\end{abstract}

\section{Introduction}
An important aspect of the dark matter problem of spiral galaxies is the
question to what extent the galaxies are dominated by dark matter even in their
inner parts, where the optically visible disks reside. The presence of dark 
matter is usually deduced
from the decompositions of the rotation curves of the galaxies. These are,
however, highly ambiguous, and `maximum' disk versus submaximal disk
decompositions of the rotation curves of spiral galaxies have been discussed at
great length in the literature. The general opinion is that the inner parts of
the galaxies are not dominated by dark matter (cf.~Bosma (1999) for a recent
review). However an increasing number of arguments have been put forward, which
challenge this conclusion:
\begin{itemize}
\item Considerations of the formation of the baryonic disks in dark halos by
dissipational collapse show that a considerable amount of dark matter is also
pulled into the inner parts during this process (Blumenthal et al.~1985).
\item Gravitational lensing of a quasar by the spiral edge--on galaxy B1600+434
indicates that its disk is submaximal (Maller et al.~2000).
\item Courteau \& Rix (1999) have used the Tully--Fisher relation to study with
a large set of galaxies the statistical correlation of the peaks of the rotation
curves of the galaxies with the radial scale lengths $h$ of the disks.
They find a dependency $v_{\rm c} \propto h^{-1/2}$, which cannot be the case,
if the galaxies were dominated by dark matter.
\item Kranz \& Rix (cf.~their contribution in this volume) have searched for the
signatures of spiral arms in the rotation curve of NGC\,4254 and conclude that
its disk is submaximal.
\end{itemize}
On the other hand, based on numerical simulations of the slowing down of
the rotation of bars in disks imbedded
in dark matter halos, Debattista \& Sellwood (1998) contend 
vociferously that galactic disks must be massive. Tremaine \& Ostriker (1999)
have made the suggestion that this problem could be possibly overcome, however, 
if the inner part of the dark halo is rotating, although Debattista \& Sellwood
(2000) claim that this would have to be at a level inconsistent with the low
rotation of the stellar halos of galaxies. In addition Weiner et al.~(2000) find
by modelling the velocity field of the barred galaxy NGC\,4123 a high
mass--to--light ratio for its bar. Thus arguments concerning the amount of dark
matter in galaxies are at present highly controversial.

The aim of this paper is to draw attention to the implications of the modelling
of the rotation curves for the internal dynamics of the disks, which are related
to the dynamical stability of the disks and the morphological appearance of the 
spiral structure of the galaxies.  

The sample of galaxies used here to study their dynamical stability has been
drawn from the list of Bottema (1993) of galaxies with measured stellar
velocity dispersions. The criteria were (a)
that the rotation curve of each galaxy, preferentially in HI, is observed,
(b) that each galaxy is so inclined that the planar velocity dispersions are
measured, but (c) that the spiral structure is clearly discernible (cf.~also
Fuchs 1999).

In section 3 density wave theory arguments will be used to descibe the
 morphology of
the spiral arms of NGC\,3223, 157, and 7083, for which NIR photometry and
rotation curves are available (Block et al.~2000). This provides again
constraints on the decomposition of the rotation curves of the galaxies.

\section{Decomposition of the rotation curves and the \protect\newline 
Q parameter as diagnostic tool}
The rotation curve of each galaxy is fitted by the superposition of
contributions due to the stellar and gaseous disks, both modelled by thin
exponential disks, in the case of NGC\,2613 a bulge, modelled by a
softened $r^{-3.5}$ density law, and the dark halo,
modelled by a quasi-isothermal sphere,
\begin{equation}
v_{\rm c}^2(R) = v_{\rm c,d}^2(R) + v_{\rm c,g}^2(R) + v_{\rm c,b}^2(R) +
v_{\rm c,h}^2(R)\,.
\end{equation}
Detailed formulae for each of the contributions in (1) can be found, for
instance, in Fuchs, M\"ollenhoff \& Heidt (1998).
The radial scale lengths of the disks, $h$, and core radii of the bulges,
$r_{\rm c,b}$, as well as the bulge to disk ratios have been adopted from
published photometry of the galaxies (cf.~Bottema (1993), Broeils (1992) and
references therein). Only in the cases of NGC\,3198 and 6503
HI data were available, which allowed the determination of the $v_{\rm c,g}$
contribution in (1). No quantitative photometry of the bulge of NGC\,7184
is available.

\begin{figure}[htb]
\begin{center}
\includegraphics[width=.99\textwidth]{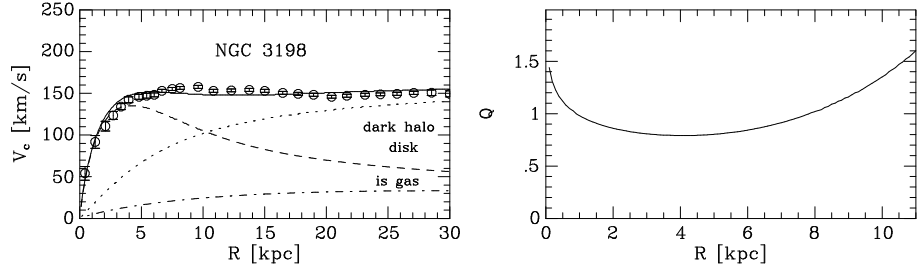}
\end{center}
\caption[]{`Maximum' disk decomposition of the rotation curve of NGC\,3198.
The contributions to the rotation curve due to the various components are
indicated. The inferred stability parameter is shown in the left panel.}
\label{fig-1}
\end{figure}

\begin{figure}[htb]
\begin{center}
\includegraphics[width=.99\textwidth]{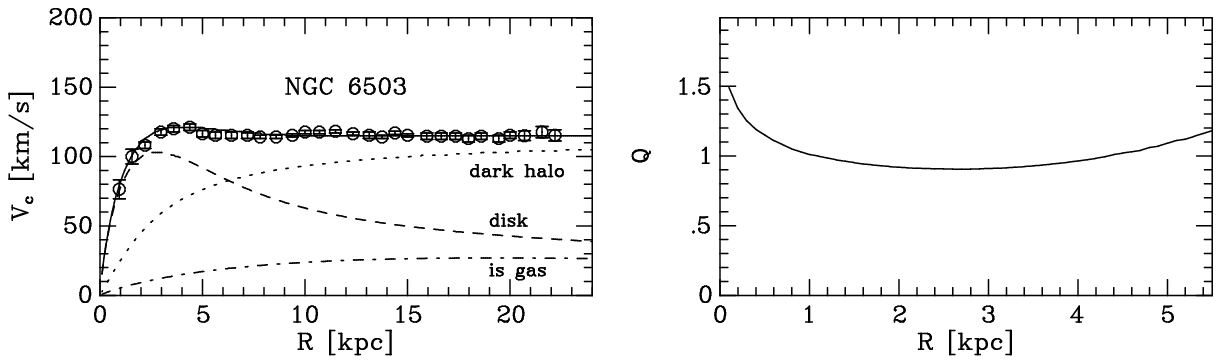}
\end{center}
\caption[]{`Maximum' disk decomposition of the rotation curve of NGC\,6503.}
\label{fig-2}
\end{figure}

\begin{figure}[htb]
\begin{center}
\includegraphics[width=.99\textwidth]{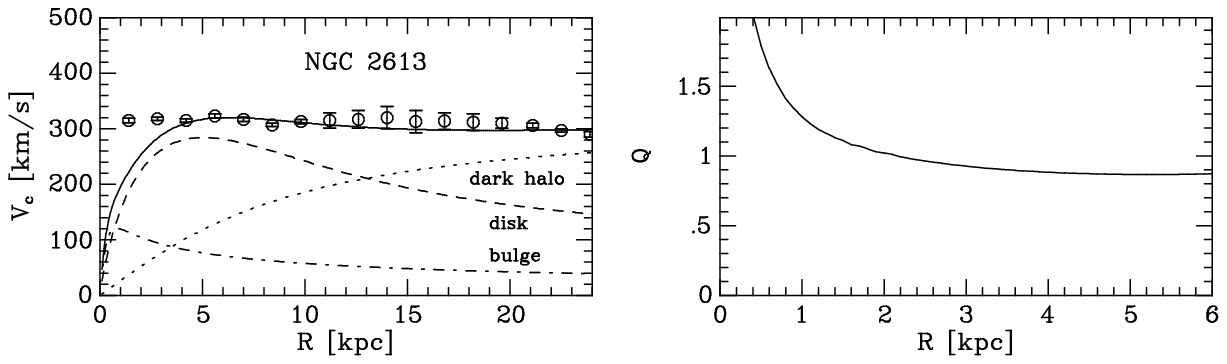}
\end{center}
\caption[]{`Maximum' disk decomposition of the rotation curve of NGC\,2613.}
\label{fig-3}
\end{figure}

\begin{figure}[htb]
\begin{center}
\includegraphics[width=.99\textwidth]{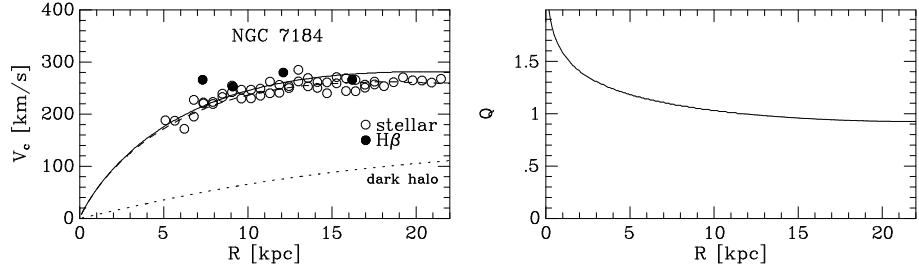}
\end{center}
\caption[]{`Maximum' disk decomposition of the rotation curve of NGC\,7184.}
\label{fig-4}
\end{figure}

The diagnostic tool, which I use to analyze the rotation curve models, is the
Toomre stability parameter of the disks, which is given by 
\begin{equation}
Q = \frac{\kappa \sigma_{\rm U}}{3.36 G \Sigma_{\rm d}}.
\end{equation}
In (2) $\kappa$ denotes the epicyclic frequency, which can be directly
derived from the rotation curve, $\sigma_{\rm U}$ the -- measured -- radial
velocity dispersion of the stars, $G$ the constant of gravitation, and
$\Sigma_{\rm d}$ the surface density of the disk, which follows from the fits
to the rotation curves. The stability parameter must lie in the range 1 $< Q
<$ 2, in order to prevent Jeans instability of the disk, on one hand, and
to allow the disks to develop spiral structures, on the other hand. 

Decompositions of the rotation curves of the galaxies, which maximise the disk
contribution in (1), are shown in Figs.~1 to 4 together with
the resulting stability parameters. As can be seen from the figures,
the $Q$ parameters are systematically close to or
even less than one. That is impossible in real galactic disks. As is well known
since the classical paper by Sellwood \& Carlberg (1984), the disks would
evolve fiercely under such conditions and heat up dynamically on short time
scales. If the model of Sellwood \& Carlberg is scaled to the dimensions of
NGC\,6503, the numerical simulations indicate that the disk would heat up
within a Gyr from $Q$ = 1 to 2.2 and any spiral structure would be
suppressed (cf.~also Fuchs \& von Linden 1998). The
amount of young stars on low velocity dispersion orbits, which would have to be
added to the disk in order to cool it dynamically back to $Q$ = 1, can be
estimated from (2). In NGC\,6503 a star formation rate of 40
$M_\odot$/pc$^2$/Gyr
would be needed, while actually a star formation rate of 1.5
$M_\odot$/pc$^2$/Gyr, as deduced from the H$_\alpha$ flux (Kennicutt et
al.~1994), is observed. Thus `maximum' disks seem to be unrealistic under this
aspect.

\begin{figure}[htb]
\begin{center}
\includegraphics[width=.99\textwidth]{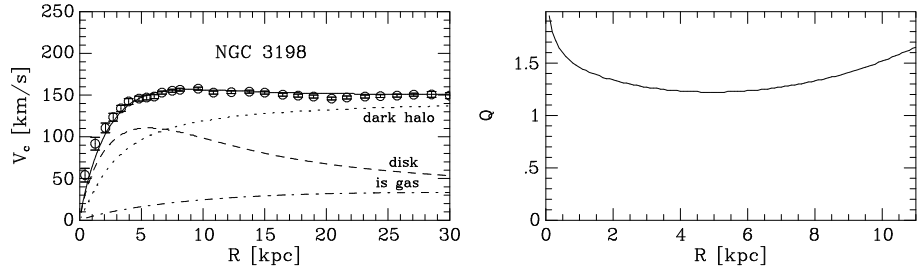}
\end{center}
\caption[]{Decomposition of the rotation curve of NGC\,3198 with a submaximal
disk.}
\label{fig-5}
\end{figure}
   
\begin{figure}[htb]
\begin{center}
\includegraphics[width=.99\textwidth]{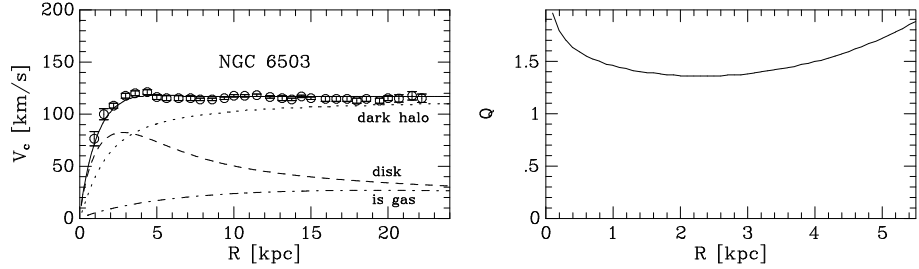}
\end{center}
\caption[]{Decomposition of the rotation curve of NGC\,6503 with a submaximal
disk.}
\label{fig-6}
\end{figure}
   
This deficiency can be remedied, if submaximal disks are
assumed. This is illustrated in Fig.~5 for NGC\,3198, where the mass-to-light
ratio of the disk has been reduced from $M/L_{\rm B}$ = 3.5
to  2.2 $M_\odot/L_{{\rm B},\odot}$. Within the
optical radius the dark halo contributes twice the mass of the disk and its
core radius is of the order of the radial scale length of the disk. As can be
seen in Fig.~5 and similarly in Fig.~6 for NGC\,6503, the $Q$
parameters lie in a more realistic range. There is, however, a caveat
about the interpretation of velocity dispersion measurements in galactic
disks that contain young stars. The problem is that stars that dominate
the spectra are relatively bright, young and have comparatively low velocity
dispersions, whereas the stars that dominate the disk mass are older and
less luminous, but have higher velocity dispersions.
For the Galactic disk this effect can be estimated quantitatively using
data from the solar neighbourhood (Jahrei{\ss}, Wielen, \& Fuchs 1998). A 
detailed analysis shows that the luminosity--weighted,
scale--height corrected radial velocity dispersion of stars in the Galactic disk
is $\sigma_{\rm U}$ = 36 km/s, which has to be compared with 44 km/s of the old
disk stars. The weight of young stars is 25\% of the total weight. In Sc
galaxies, which are bluer than the Galaxy with an averaged $\langle B-V \rangle$
of 0.66 mag, this might be shifted even more towards young stars. On the other
hand, Sc galaxies are more gas rich, which has a destabilizing effect. Taken all
together, the $Q$ argument seems to be quite robust.
\begin{figure}[htb]
\begin{center}
\end{center}
\caption[]{$K'$ image of NGC\,3223 (reproduced from Grosb{\o}l \& Patsis 1998).}
\label{fig-7}
\end{figure}

\section{Spiral density wave theory constraints on the decomposition of  
rotation curves}

The morphology of spiral galaxies is described theoretically by the density
wave theory of galactic spiral arms. One of the predictions of the
theory is that spiral density waves with a circumferential wave length of
\begin{equation}
\lambda = X\left(\frac{\D v_{\rm c}}{\D R}\right) \cdot \lambda_{\rm crit}
= X\left(\frac{\D v_{\rm c}}{\D R}\right) \cdot \frac{4 \pi^2 G 
\Sigma_{\rm d}}{\kappa^2}
\end{equation}
have the largest growth rate and will dominate the morphological appearance
of the disk (Toomre 1981).

\begin{figure}[htb]
\begin{center}
\includegraphics[width=.36\textwidth]{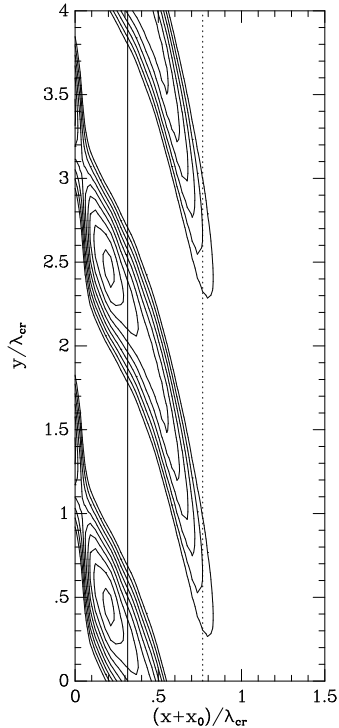}
\end{center}
\caption[]{Theoretical model of the spiral arms of NGC\,3223 (Block et
al.~2000). The $x$--axis points radially outwards and the $y$--axis in the
tangential direction. Length unit is $\lambda_{\rm crit}$.
The contours are given in arbitrary units. The disk is bounded at the left side
($x=-x_0$) by the massive bulge of the galaxy.} 
\label{fig-8}
\end{figure}
\noindent
The value of the $X$ parameter is about 2 in the case of a flat rotation curve,
but less for rising rotation curves (Athanassoula et al.~1987, Fuchs 1999). The
expected number of spiral arms is given by
\begin{equation}
m = \frac{2 \pi R}{\lambda}\,.
\end{equation}
Block et al.~(2000) have presented theoretical models based on the work of Fuchs
(2001, in preparation) for the spiral arms of the galaxies NGC\,3223, 157, and
7083, for which NIR photometry and rotation curves are available. The epicyclic
frequency and the value of the $X$ parameter have been obtained directly from
the observed rotation curves for each galaxy. The spiral arms of all three 
galaxies are very regular and clearly defined in the NIR (see Block et
al.~2000 for images). Equations (3) and (4) can be then used to determine
the surface densities of the disks, modelled again by thin exponential disks.
NGC\,3223 is a good example to illustrate the constraint on the disk mass
obtained this way from the morphology of the galaxy in some more detail. In 
Fig.~7 a NIR image is reproduced and in Fig.~8 the theoretical model of the
spiral arms of the galaxy is shown, where the same value of Oort's
constant $A/\Omega_0$ as in the optical part of NGC\,3223 has been adopted.
The theoretical model, which has been 
calculated in rectangular coordinates, has to be imagined to be folded around
the galactic center of the galaxy. Since NGC\,3223 has two well defined spiral 
arms, this constrains the critical wave length $\lambda_{\rm crit}$ according to
(3) and (4). This, in turn, determines the surface density of the disk. Together
with the
radial scale length of the disk measured by Grosb{\o}l and Patsis (1998) this
allows finally to calculate the disk rotation curve (cf.~(1)) shown in Fig.~9.
Obviously the disk of NGC\,3223 is submaximal, and the same
is also found for the two other galaxies.

\begin{figure}[htb]
\begin{center}
\includegraphics[width=.6\textwidth]{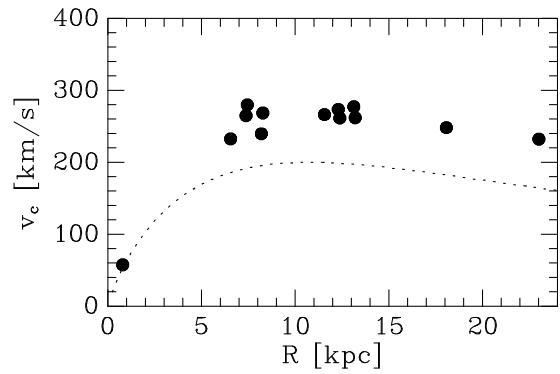}
\end{center}
\caption[]{Rotation curve of NGC\,3223. The observed data are indicated by the
symbols and the disk contribution to the rotation curve is shown by the
dotted line.} \label{fig-9}

\end{figure}

%INDEX%%%%%%%%%%%%%%%%%%%%%%%%%%%%%%%%%%%%%%%%%%%%%%%%%%%%%%%%%%%%%%%
% Please check with the editor of your book whether he plans to
% include a "mutual" subject index - if so, please code your entries
% in the standard syntax. For your own purposes you may print your
% "personal" index by using the following commands:
%
%\clearpage
%\addcontentsline{toc}{section}{Index}
%\flushbottom
%\printindex
%%%%%%%%%%%%%%%%%%%%%%%%%%%%%%%%%%%%%%%%%%%%%%%%%%%%%%%%%%%%%%%%%%%%%

\end{document}